Switchbacks in the solar magnetic field:
their evolution, their content, and their effects on the plasma

SHORT TITLE: SWITCHBACK EVOLUTION, CONTENTS AND EFFECTS ON THE PLASMA

**F.S. Mozer**[1], O.V. Agapitov[1], S.D. Bale[1], J.W. Bonnell[1], T. Case[4], C.C. Chaston[1], D.W. Curtis[1], T. Dudok de Wit[6], K. Goetz[2], K.A. Goodrich[1], P.R. Harvey[1], J.C. Kasper[5], K.E. Korreck[4], V. Krasnoselskikh[6], D.E. Larson[1], R. Livi[1], R.J. MacDowall[7], D. Malaspina[3], M. Pulupa[1], M. Stevens[4], P.L Whittlesey[1], J.R. Wygant[2]

(1) University of California, Space Science Laboratory, Berkeley, CA, United States
(2) University of Minnesota, Minneapolis, MN, United States
(3) University of Colorado, Boulder, Laboratory for Atmospheric and Space Physics, Boulder, CO, United States
(4) Smithsonian Astrophysical Observatory, Center For Astrophysics, Cambridge, MA, United States
(5) University of Michigan, Ann Arbor, MI, United States
(6) LPC2E, CNRS and University of Orléans, Orléans, France
(7) NASA/Goddard Space Flight Center, Greenbelt, United States

Switchbacks (rotations of the magnetic field) are observed on the Parker Solar Probe. Their evolution, content, and plasma effects are studied in this paper. The solar wind does not receive a net acceleration from switchbacks that it encountered upstream of the observation point. The typical switchback rotation angle increased with radial distance. Significant Poynting fluxes existed inside, but not outside, switchbacks and they are related to the increased **E**X**B**/$B^2$ flow caused by the magnetic field rotating to become more perpendicular to the flow direction. (Outside the switchbacks, the magnetic field and solar wind flow were generally radial.) The solar wind flow inside switchbacks was faster than that outside due to the frozen-in ions moving with the magnetic structure at the Alfven speed. This energy gain results from the divergence of the Poynting flux from outside to inside the switchback, which produces a loss of electromagnetic energy on switchback entry and recovery of that energy on exit, with the lost energy appearing in the plasma flow. Switchbacks contain 0.3-10 Hz waves that may result from currents and the Kelvin-Helmholtz instability that occurs at the switchback boundaries. These waves may combine with lower frequency MHD waves to heat the plasma. The radial decreases of the Poynting flux and solar wind speed inside switchbacks are due to a geometrical effect.

An interesting result of the Parker Solar Probe mission has been the observation of switchbacks in the solar magnetic field. Switchbacks are seconds to minutes duration rotations of the magnetic field through angles that can be larger than 90°. It is the purpose of this paper to discuss the evolution of switchbacks, their wave and plasma content, and their relationship to the acceleration of solar wind ions.

The Parker Solar Probe and its instruments are described elsewhere [Bale et al, 2016; Fox et al, 2016; Kasper et al, 2016]. It is in a solar orbit with its first perihelion at 35 solar radii ($R\odot$) occurring on 2018 November 5 and its second perihelion at 35 $R\odot$ occurring on 2019 April 5. The coordinate system used in the following discussion is tied to the spacecraft, has X perpendicular to the Sun-spacecraft line, in the ecliptic plane, and pointing in the direction of solar rotation (against the ram direction), Y perpendicular to the ecliptic plane, pointing southward, and Z pointing sunward. Figure 1 presents an overview of 10 days of fields and plasma data collected around the April perihelion. Figure 1a gives the Z-component of the magnetic field, which contains many spikes, which are the switchbacks [Bale et al, 2019, Kasper et al, 2019]. Such structures were known prior to the Parker Solar Probe launch [Yamauchi et al, 2004; Landi et al, 2005, 2006; Suess, 2007; Neugebauer et al, 2013; Matteini et al, 2014; Borovsky, 2016; Horbury et al, 2018]. That the switchbacks are primarily rotations of the magnetic field is shown by the fact that the total magnetic field in Figure 1e was nearly devoid of variation as the switchbacks occurred.

To study the evolution of the switchbacks in time, Figure 2 presents the magnetic field rotation angles as functions of time for days near the ~35 (actually 36.0) $R\odot$ perihelion and, five days earlier, at ~50 (actually 48.2) $R\odot$. To the eye, there appear to be more switchbacks at the outer radius and they involved much larger field rotations than those at perihelion. This result is made quantitative in Figure 3, which gives the number of events having rotations greater than $30^0$, $60^0$ and $90^0$ on the two days. That there were many more switchbacks at the outer radius than at perihelion suggests that even fewer and smaller switchbacks occur closer to the Sun. Related statistical analyses on switchback occurrence are discussed in an accompanying article (Dudek de Wit et al, 2019).

As seen in Figure 1, switchbacks contain large Poynting fluxes (Figure 1b, which gives **E**X**B**/$\mu_0$, where **E** and **B** are the electric and magnetic field), energized ion kinetic energy (Figure 1c, which gives $0.5mv^2$, where v is the plasma bulk flow), and hotter radial plasma (Figure 1d, which is $0.5mw^2$, where w is the radial plasma thermal velocity). The Poynting flux of Figure 1b measures the Z-component of the electromagnetic energy, which was negative, meaning that the electromagnetic energy flowed away from the Sun, and it was spiky because the enhanced Poynting flux was embedded in magnetic field switchbacks. A striking feature of the Poynting flux is that it was large near perihelion and an order-of-magnitude smaller at 50 $R\odot$, both before and after perihelion. This same spatial dependence was observed over the same radial distances on the first perihelion pass, so these four observations are assumed to represent the spatial distribution of the Poynting flux. The proton kinetic energy of Figure 1c increased from 500 eV/particle to nearly 2000 eV/particle and the ion temperature of Figure 1d increased from 20 to 60 eV in regions of enhanced Poynting flux in switchbacks near perihelion.

Examples of switchbacks that illustrate these enhancements, at 35 and 50 solar radii in Figure 1, are given in Figure 4. Figures 4c and 4k give the radial components of the magnetic field whose value near perihelion at 35 $R_\odot$ changed from +100 to -90 nT while, at 50 $R_\odot$, the field change was from 50 to 15 nT. These changes are characteristic of decreases of Poynting flux with distance and they identify these events as large amplitude switchbacks. Within the switchbacks, the largest Poynting flux was 8000 µW/m$^2$ at 35 $R_\odot$ (Figure 4d) and 1000 µW/m$^2$ at 50 $R_\odot$ (Figure 4l). The ion kinetic energy increased from 1000 to 3000 eV at 35 $R_\odot$ (Figure 4e) and from 500 to 600 eV at 50 $R_\odot$ (Figure 4m), while the radial ion temperature increased from 20 to 60 eV at perihelion (Figure 4f) and from 15 to 20 eV at 50 $R_\odot$. During these times the plasma densities (Figures 4g and 4o) and total magnetic fields Figures 4h and 4p) remained roughly constant. These variations are characteristic of large switchbacks seen at the two radial distances.

Figures 4a and 4i, and 4b and 4j present spectra of the perpendicular electric and magnetic fields, respectively. They show that the main power in the waves was at frequencies more than an order-of-magnitude below the ~1.5 Hz ion gyrofrequency or the Doppler-shifted ion inertial scale fluctuations (tens of Hz at these densities and solar wind speeds). They also show that such waves had much larger amplitudes at 35 $R_\odot$ than at 50 $R_\odot$. This suggests that the waves were in the long-wavelength Alfven mode. It is noted that there was also wave power at 0.01-10 Hz inside the switchbacks and these waves will be discussed below.

The next topic will be investigation of the source and effects of the enhanced Poynting flux, ion bulk flow, and ion temperature. Figure 5 presents plots of the Poynting flux as functions of the magnetic field rotation angle at the two distances. It shows that the magnitude of the Poynting flux increased rapidly with the rotation angle of the switchbacks and was an order-of-magnitude larger at perihelion than at 50 $R_\odot$. That the large Poynting fluxes did not propagate and were not present outside switchbacks is shown by the absence of large Poynting fluxes in the dashed rectangles in Figure 5. These rectangles are outside of switchbacks because they are located at small rotation angles. Their area is where large Poynting fluxes would appear if such Poynting fluxes were created in upstream switchbacks and if they propagated to the observed locations, outside the switchbacks. At both radial distances, there are not significant Poynting fluxes in the dashed rectangles. Thus, one may conclude that the Poynting fluxes are confined to the interiors of switchbacks and they neither propagate nor are they generated outside of switchbacks.

The confinement of the Poynting flux to the interior of switchbacks is understood by realizing that, outside of the switchbacks, the radial solar wind flows mostly parallel to the magnetic field. However, as the magnetic field rotates, more of the radial solar wind flow becomes perpendicular to the magnetic field, so it becomes an **E**X**B**/B$^2$ flow. Because the Poynting flux is **E**X**B**/µ$_0$, it must also increase along with **E**X**B**/B$^2$ as the magnetic field rotates. The more the field rotates, the more the solar wind flow becomes perpendicular to the magnetic field, so the Poynting flux must increase as B rotates. As the rotation angle exceeds 90$^0$, **E**X**B**/B$^2$ decreases and the Poynting flux decreases, as is seen in Figure 5. In addition to explaining the observations of Figure 5, this analysis suggests that there is little or no long term exchange of energy between the Poynting flux and the plasma bulk flow.

One may also understand from this analysis why the magnitude of the Poynting flux decreases with radial distance, as is evident in Figures 1b, 4, and 5. Because the solar wind speed is roughly constant between 35 and 50 $R\odot$, **E**X**B**/$B^2$ at a fixed rotation angle is also constant. Thus, **E**X**B**/$\mu_0$ is proportional to $B^2$, or $1/R^4$, where R is the radial distance from the Sun. This is a factor of about four between 35 and 50 $R\odot$, which is in general agreement with the radial decrese of Poynting flux in Figures 1b, 4 and 5. It also suggests that the Poynting flux should increase rapidly at lower altitudes. However, because the magnetic field rotation likely decreases rapidly at lower altitudes (as suggested by Figure 2), this Poynting flux increase may not occur.

The variation of the ion flow velocity in switchbacks at 35 and 50 $R\odot$ is illustrated in Figure 6, [also noted at larger solar radii by Matteini et al, 2014] in which the black dots are the total proton bulk speed and the red dots are the total magnetic field. That the amplitudes of the red curves are independent of the magnetic field rotation angle shows that the field change was truly a rotation of the field.

Energized ions occur in switchbacks, as shown in Figures 1c, 4, and 6. That such ions are confined to the interiors of switchbacks is also evident because there are no such ions inside the dashed rectangles of Figure 6, which are the locations where more energetic ions, created in upstream switchbacks, would appear in local regions that are outside of switchbacks.

To study the mechanism behind the increase of the ion velocity in switchbacks, the locally parallel and perpendicular velocities as functions of the magnetic field rotation are plotted in Figure 7 for the 24 hours when the Parker Solar Probe was at its 35 $R\odot$ perihelion. As the magnetic field rotated to $90^0$ inside switchbacks, the parallel (radial) velocity decreased to zero and the perpendicular (radial) velocity increased. At $90^0$ rotation, the perpendicular speed was about 600 km/s, while, at $0^0$ rotation, the parallel speed was 300 km/s. This shows that there was a net increase of total velocity as the magnetic field rotated. The perpendicular speed increased because it included the Alfven speed associated with the magnetic field perturbation. Thus, the ion velocity increase inside switchbacks was caused by the frozen-in ions moving with the magnetic field as it varied in the Alfven wave. The plots in figure 7 are curved. This is because they are proportional to the sine and cosine of the rotation angle, respectively.

The ion velocity in Figure 6a changed from about 400 km/s to 600 km/s as the magnetic field rotated from $0^0$ to $90^0$ at perihelion. AT 50 $R\odot$, the change in Figure 6b was from 350 km/s to 450 km/s. That the velocity change was a factor of about two greater at 35 $R\odot$ can be, at least partially, understood by the 1/R decrease of the Alfven speed between the two points, which is a factor of about 1.5.

A question arises as to where the ions got their additional energy inside the switchbacks. The divergence of the Poynting flux, $(\nabla \cdot \mathbf{S})$, gives the change of electromagnetic energy across the divergence region according to Poynting's theorem. There is no such divergence from one side of the switchback to the other so there is no net electromagnetic energy gain or loss and, therefore, no solar wind energy change due to the switchback. However, there is a net divergence of the Poynting flux from outside to inside the switchback so there must be a loss of electromagnetic energy on entering the switchback and a nearly equal gain of electromagnetic energy on its exit. These divergences result in the ions inside the switchback having a greater kinetic energy than those outside. This solar wind energy gain inside a typical switchback may be estimated by assuming that the solar wind velocity outside the switchback is 400 km/sec, the density is 80 cm$^{-3}$, and the Poynting flux inside the switchback is 6000 µW/m$^2$. For these values, the solar wind velocity gain is 200 km/sec and the energy changes from 830 eV outside the switchback to 1900 eV inside. These results are in good agreement with the observed changes of velocity and kinetic energy in Figures 6 and 7.

The increase of ion temperature inside switchbacks at the two radial distances is illustrated in Figure 8. This may be due to one or both of the following reasons. First, the temperature measurement is made only for the plasma component along the line of sight to the sun. As the switchback rotates, the measured quantity changes from being the parallel temperature, $T_{par}$, to the perpendicular temperature, $T_{perp}$. Thus, if $T_{perp}/T_{par} > 1$, the measured temperature would increase with the switchback rotation angle. Second, because of the waves discussed below, it is possible that the plasma is heated in the switchbacks. An ongoing analysis will attempt to distinguish between these two explanations. It is noted that the ion temperature at the outer radius is less than that at 35 $R\odot$. This is because the ions cool as they and the magnetic field expand.

Heating cannot be produced by long wavelength, low frequency Alfvenic turbulence alone, so additional waves must be associated with switchbacks if there is heating. Such waves are present in the wavelet spectra of Figures 4a and 4b where the wave intensity at ~0.3-10 Hz is enhanced at the switchback boundaries and inside the switchbacks. Figure 9 presents expanded views at 35 and 50 $R\odot$ of the ~0.3-10 Hz waves and turbulence in the data of Figure 4. Figures 9a and 9d present the three components of the magnetic field at the two locations with the changes of $B_z$ indicating the switchback boundaries. Figures 9b and 9e present the fluctuations in the magnetic field with frequencies greater than 0.1 Hz and figures 9c and 9f give their spectra at ~0.3-10 Hz. The amplitude of the fluctuations at 35 $R\odot$ was about a factor of three greater than that at 50 $R\odot$.

Peaks in the spectra of Figures 4 and 9 occurred at the boundaries of the switchbacks, which suggests that waves were generated at these boundaries. Because the ion velocity inside the switchbacks was greater than that outside, it is feasible that the boundary was Kelvin-Helmholtz (KH) unstable and that this provided the source of these surface waves. To test this possibility, the fields were examined in the minimum variance coordinate system of the magnetic field at each switchback boundary. The velocities at the beginning of the 35 $R_\odot$ switchback, near 16:43 UT and at its end, near 16:44 UT, show significant velocity shears (about 260 kms-1 and 150 kms-1 respectively) with almost zero velocity along the normal. These boundaries would be unstable to the KH instability if the following inequality is satisfied [Miura, 2003; Parks, 2004]

$$\rho_1\rho_2[(\mathbf{v_1}-\mathbf{v_2})\cdot\mathbf{k}]^2 > (\rho_1+\rho_2)[(\mathbf{B_1}\cdot\mathbf{k})^2 + (\mathbf{B_2}\cdot\mathbf{k})^2]/\mu_o \qquad (1)$$

where the subscripts refer to the two sides of a boundary, $\rho$ is the mass density, $\mathbf{k}$ is the wave vector, $\mathbf{v}$ is the velocity, and $\mathbf{B}$ is the magnetic field. For the wave normal vector directed along the velocity shear, this inequality is satisfied by a factor of about two at the leading edge and 1.4 at the trailing edge of the switchback at 35 $R_\odot$. Applying the same criteria to the data at 50 $R_\odot$ shows that this crossing was marginally stable.

There are two types of perturbations associated with the solar wind switchbacks: the large-scale Alfvenic perturbation associated with the rotation of the magnetic field and the small-scale waves inside the structure that are generated at the structure's boundaries [Krasnoselskikh, et al, 2019]. The smaller-scale ~0.3-10 Hz waves are MHD modes because their Doppler shifted frequencies are below the ion cyclotron frequency of a few Hz. These perturbations affect the plasma in two ways, first by the large-scale reversible enhancement of the plasma bulk velocity in the switchback, and second, by predominantly pitch angle scattering of the ions by the small-scale waves. This scattering of large bulk-velocity protons may spread part of the bulk flow kinetic energy and lead to heating of protons inside the switchback. Thus, the combined interactions of these different scale perturbations with protons may guide the transfer of switchback energy into thermal energy during the switchback propagation.

**SUMMARY**

The switchback evolution in time and space has been studied with the conclusion that both the number of switchbacks per unit time and the magnetic field rotation inside them increase with distance from the Sun. The switchbacks contained enhanced solar wind bulk flow, Poynting flux and thermal energy. The Poynting flux increased in switchbacks as a consequence of the rotation of the magnetic field such that the bulk flow changed from being parallel to the magnetic field to having a component perpendicular to the field. This perpendicular flow is $\mathbf{E}\mathbf{X}\mathbf{B}/B^2$. Hence, $\mathbf{E}\mathbf{X}\mathbf{B}/\mu_0$, the Poynting flux, also increased as the magnetic field rotated. For a fixed angular rotation, the Poynting flux varies with the radial distance from the Sun, as $1/R^7$. Thus, it is nearly an order-of-magnitude smaller at 50 solar radii than at 35 solar radii, as is observed. This radial dependence suggests the possibility that the Poynting flux is huge at smaller radii. This may be unlikely because the rotation angle also decreases at smaller radii. The Poynting flux outside switchbacks is essentially zero because the bulk flow is almost parallel to B. Thus, the energy associated with the Poynting flux does not accelerate the bulk ion flow.

Even so, the bulk ion flow inside switchbacks was observed to increase relative to that outside. This is because the ions are tied to the oscillating magnetic field, so they gain an additional velocity proportional to the Alfven speed, which they lose after the switchback passes by. This energy comes from the divergence of the Poynting flux of opposite signs on entering and exiting the switchback. Because the Alfven speed is proportional to 1/R, the enhanced ion speed inside switchbacks is smaller at 50 solar radii than at 35 solar radii.

Wave power is observed at ~0.3-10 Hz at the boundaries and inside switchbacks. These waves may result from large velocity shears at the switchback boundaries, which cause the boundaries to be Kelvin-Helmholtz unstable.

**ACKNOWLEDGEMENTS**
This work was supported by NASA contract NNN06AA01C. The authors acknowledge the extraordinary contributions of the Parker Solar Probe spacecraft engineering team at the Applied Physics Laboratory at Johns Hopkins University.

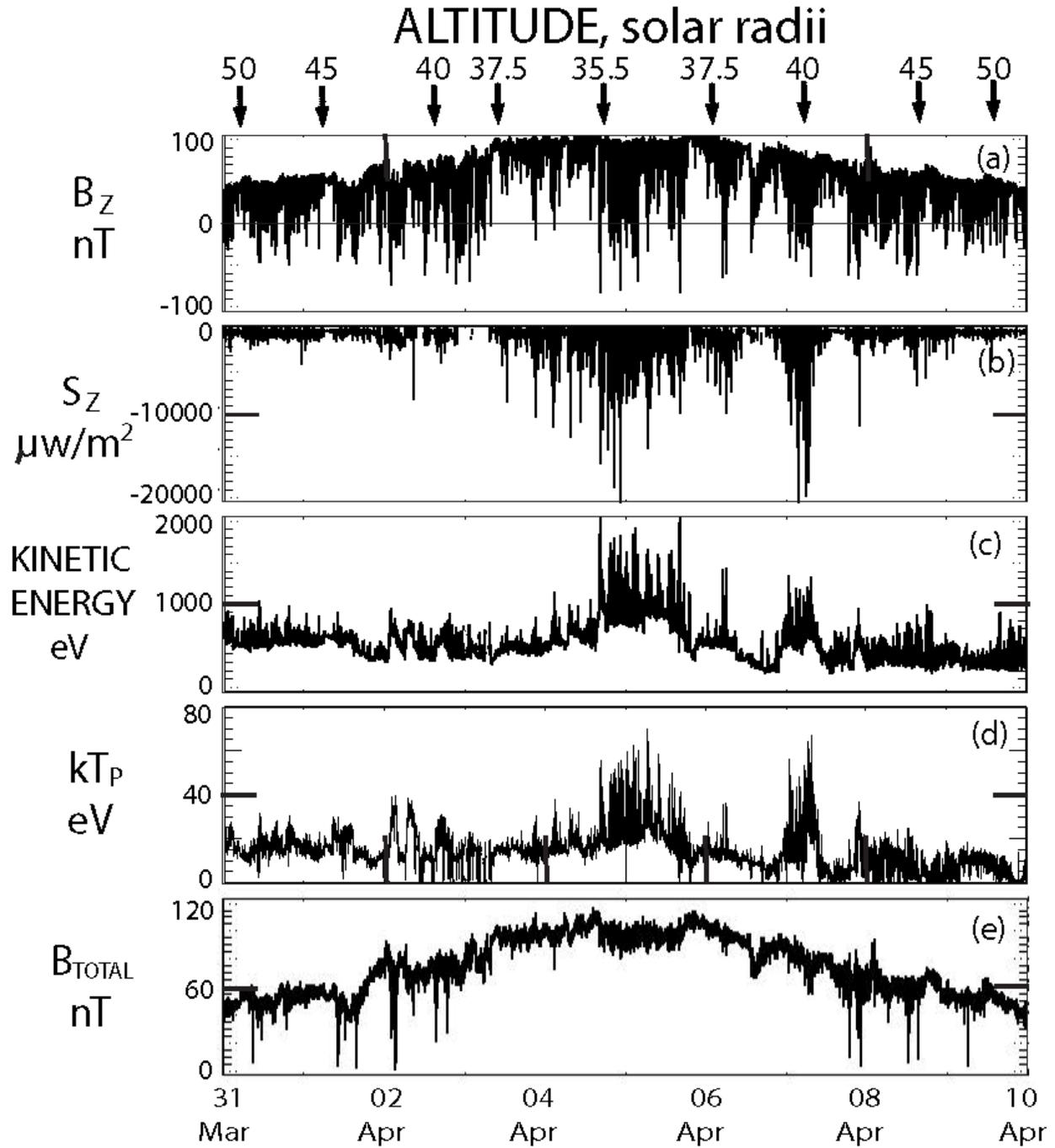

Figure 1. Field and particle measurements during ten days of spacecraft passage through perihelion on 2019 April 4. Switchbacks in the magnetic field (Figure 1a) are associated with enhancements of the Poynting flux (Figure 1b), and increases of the solar wind kinetic energy and apparent temperature (Figures 1c and 1d).

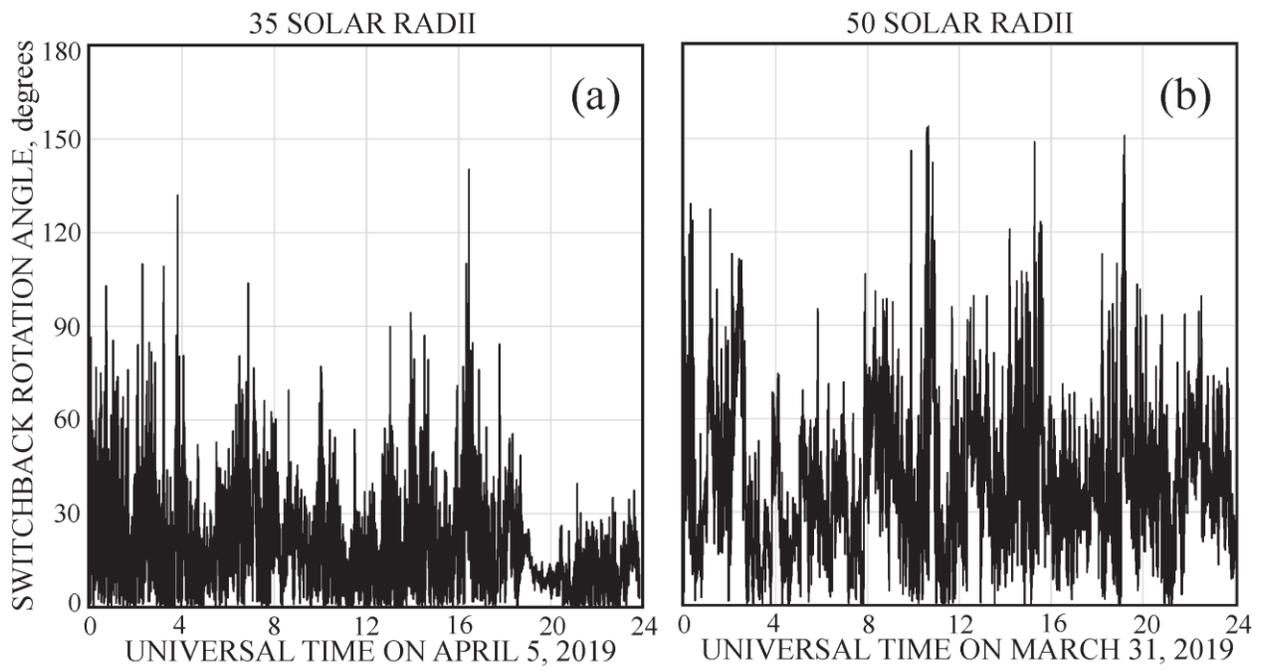

Figure 2. The switchback rotation angle as a function of time during one-day intervals at the perihelion distance of 35 solar radii and at 50 solar radii.

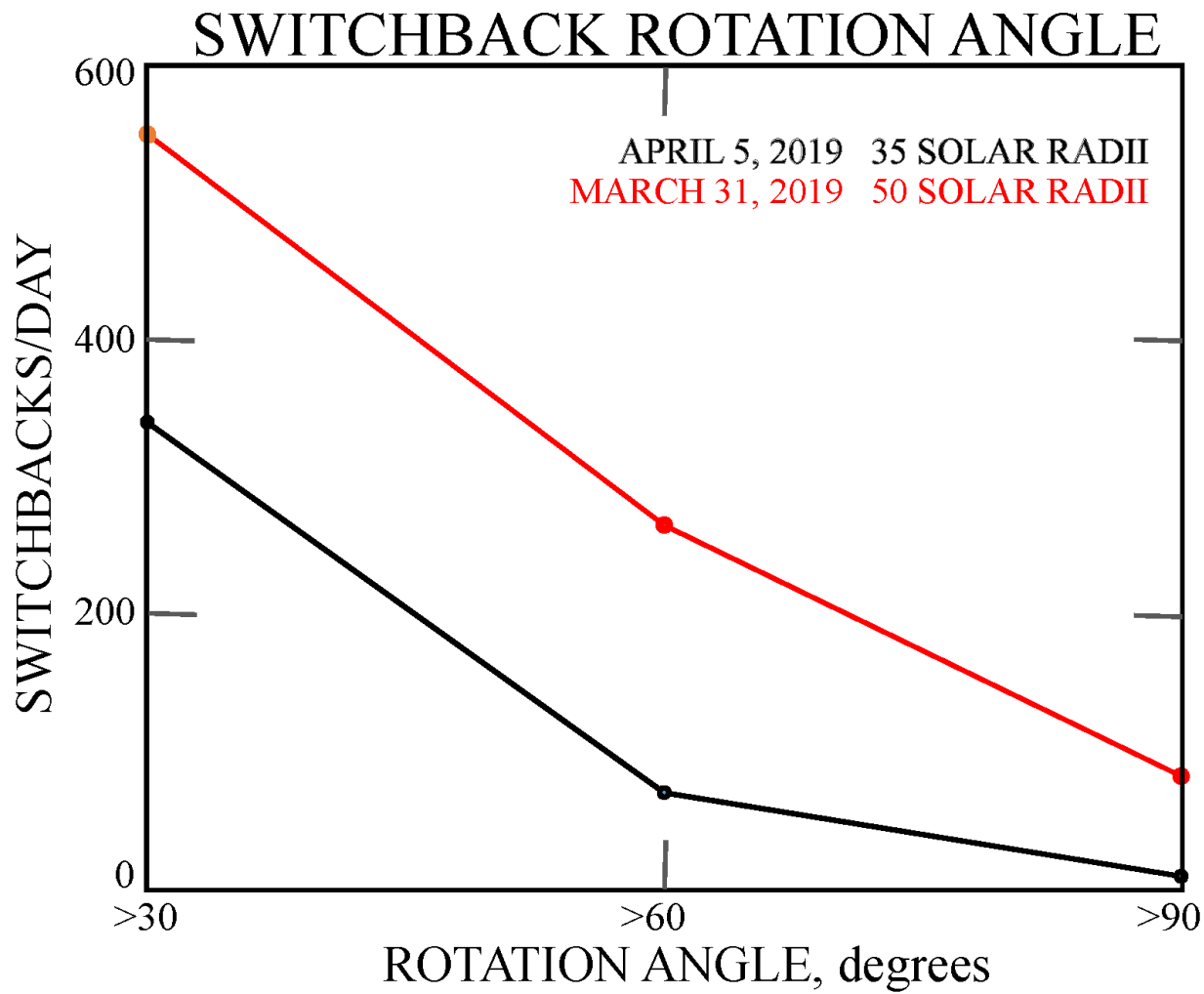

Figure 3. The number of switchbacks per day versus their rotation angles at 35 and 50 solar radii. The increased number of large angle switchbacks at further radial distances is illustrated.

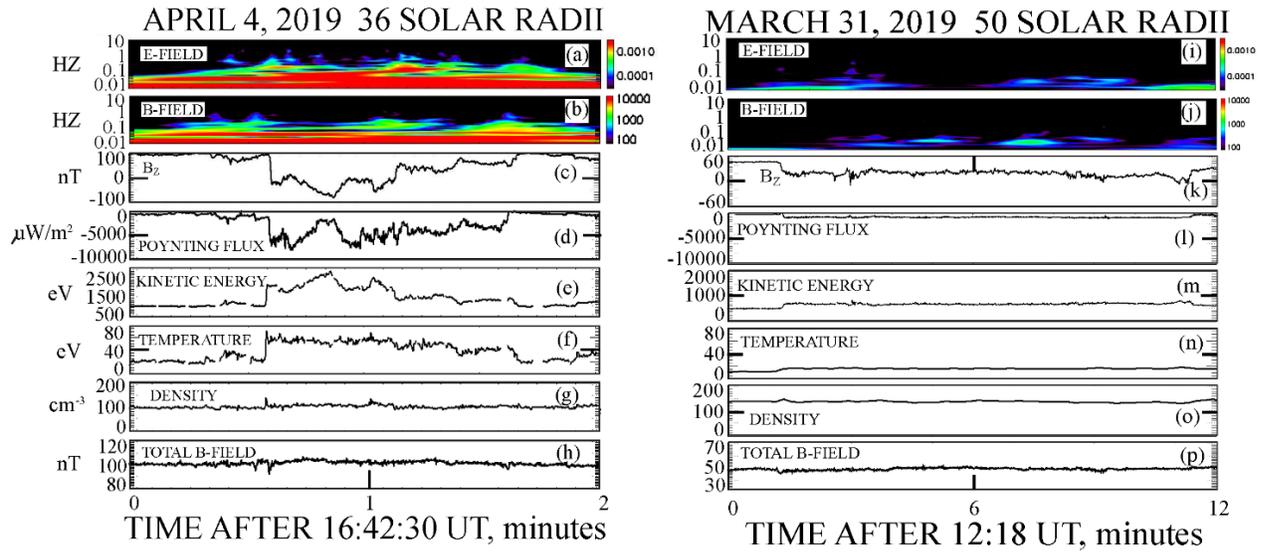

Figure 4. The wavelet spectra of the perpendicular electric field at 35 $R_\odot$ (Figure 4a) and 50 $R_\odot$ (Figure 4i), magnetic field wavelet spectra (Figure 4b and 4j), and plasma and field parameters at typical switchbacks. The wave power spectra peaked at the abrupt changes of the magnetic field. Because the wave power was at low frequencies and the total magnetic field did not change (figures 4h and 4p), the waves were Alfven mode waves. Also note the significantly greater variations of all parameters at 35 $R_\odot$ as compared to 50 $R_\odot$.

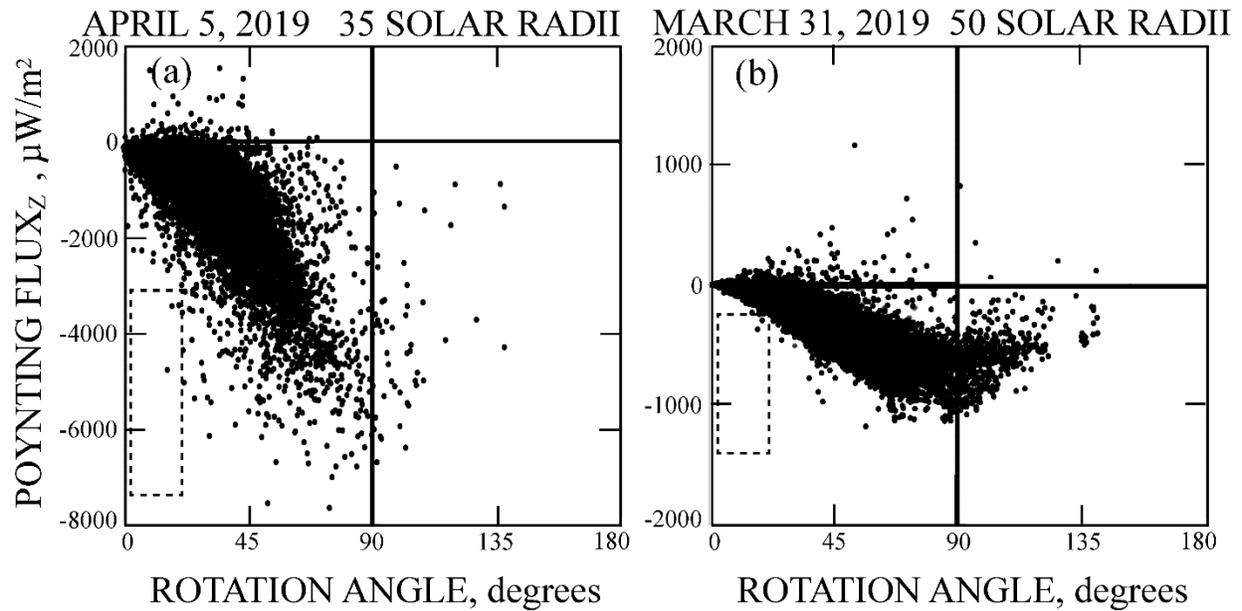

Figure 5. The radial Poynting flux as a function of the rotation angle of switchbacks at 35 and 50 $R_\odot$. The angular range of switchback rotations was greater at 50 $R_\odot$ and the Poynting flux at $90^0$ was an order-of-magnitude larger at 35 $R_\odot$ than at 50 $R_\odot$. Also note that the Poynting flux appears to be maximum at $90^0$ and to decrease for larger rotations. The Poynting flux was not significant outside switchbacks, i.e., inside the dashed rectangles.

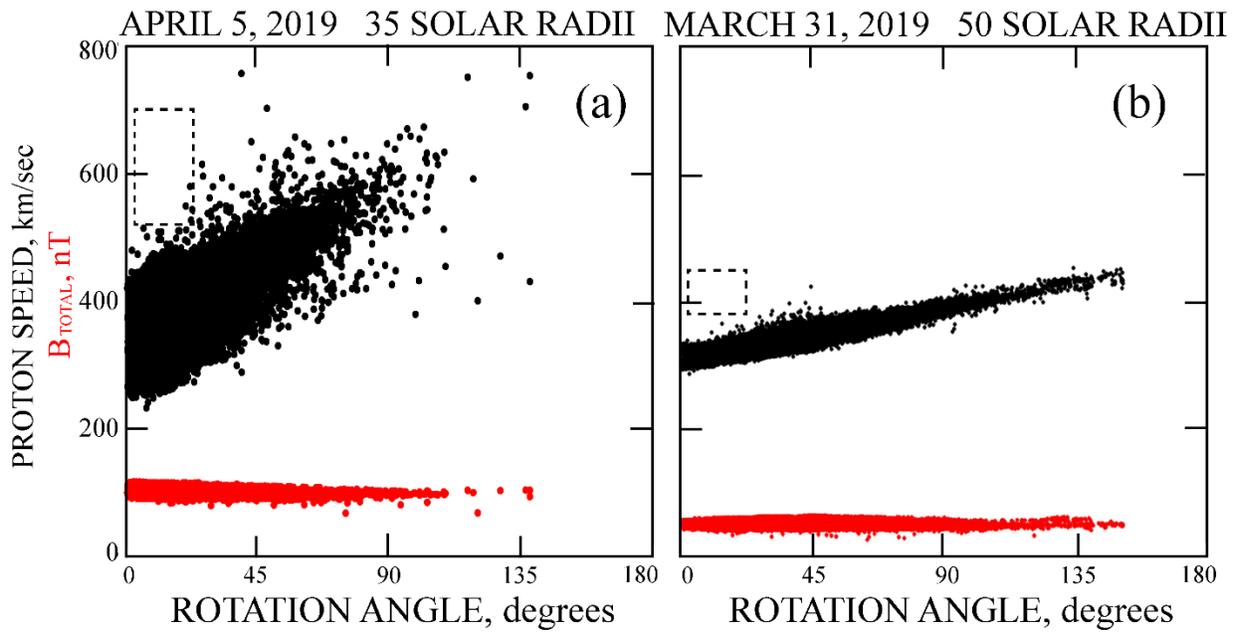

Figure 6. The proton bulk speed and the total magnetic field as functions of the switchback rotation angle at 35 and 50 solar radii. Because the total magnetic field did not vary with the rotation angle, the magnetic field variations must have been true rotations. The proton bulk sped is not enhanced outside of the switchbacks, i.e., inside the dashed rectangles.

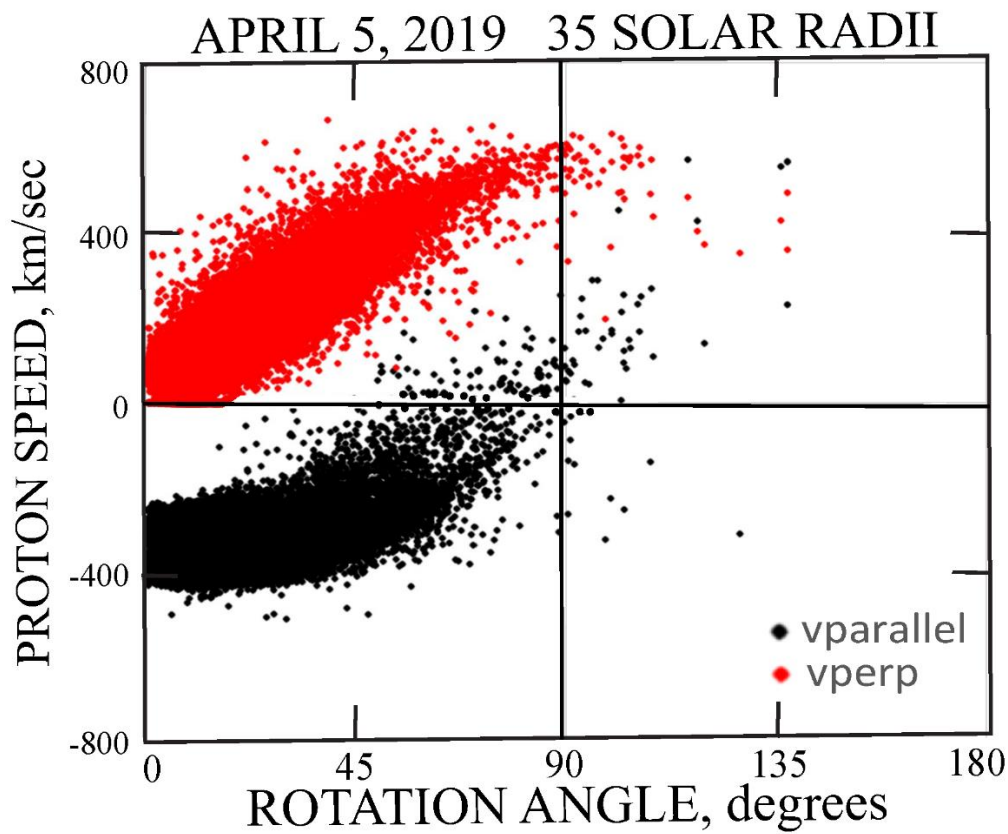

Figure 7. The local perpendicular and parallel components of the solar wind flow speed as functions of the magnetic field rotation angle. For zero rotation, the proton speed is parallel to the radial magnetic field while at 90 degree switchbacks, the radial flow is perpendicular to the local magnetic field and it is larger than the flow speed outside of switchbacks (i.e., at zero degree rotations).

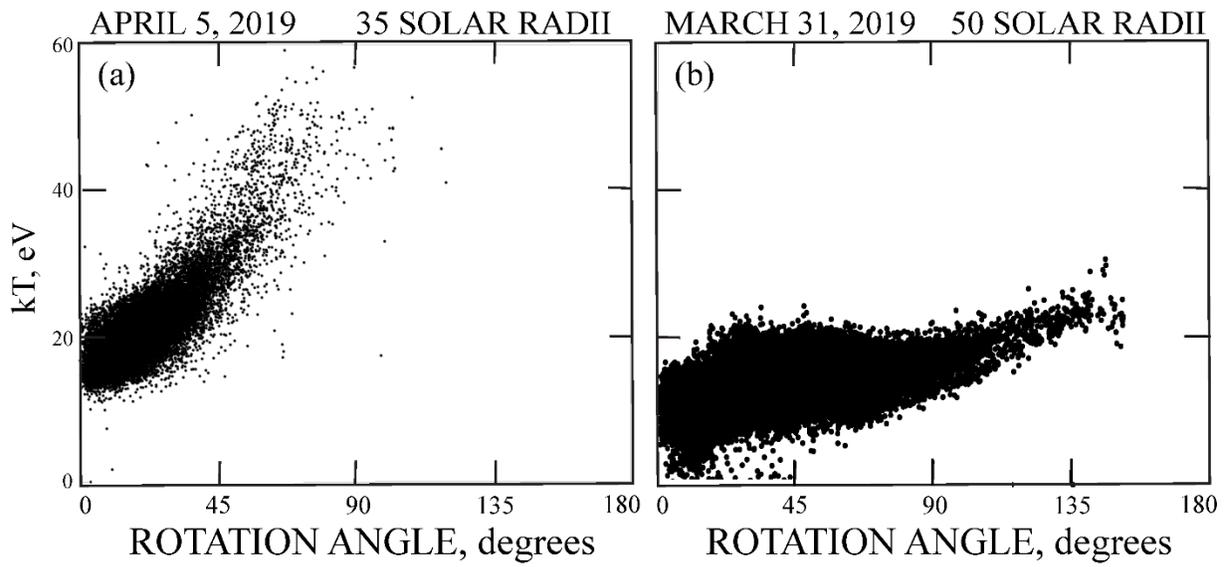
Figure 8. The ion temperature as a function of the switchback rotation angle.

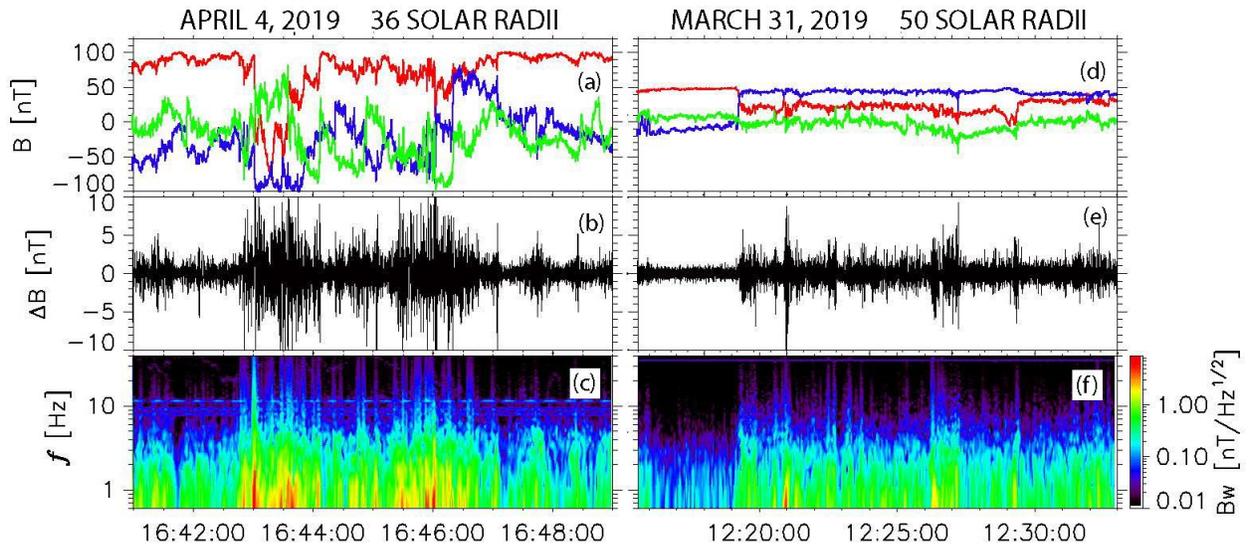

Figure 9. Magnetic field (Figures 9a and 9d), magnetic fluctuations (Figures 9b and 9e), and the magnetic field spectra (Figures 9c and 9f) during the switchbacks illustrated in Figure 4.